\documentclass[aps,superscriptaddress,showpacs]{revtex4-1}

\usepackage{graphics}
\usepackage{epsfig}
\usepackage{dcolumn}
\usepackage{epsfig}
\usepackage{mathrsfs}
\usepackage{amsfonts}
\usepackage{amsmath}
\usepackage{amssymb}
\newcommand{\ii}{\raisebox{-0.4\height}{\includegraphics[height=0.5cm]{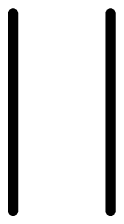}}}
\newcommand{\D}{\raisebox{-0.4\height}{\includegraphics[height=0.3cm]{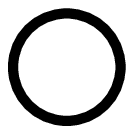}}}
\newcommand{\U}{\raisebox{-0.4\height}{\includegraphics[height=0.5cm]{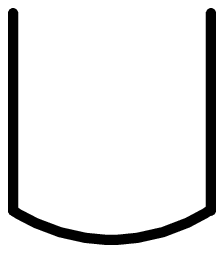}}}
\newcommand{\Uij}{\raisebox{-0.4\height}{\includegraphics[height=0.7cm]{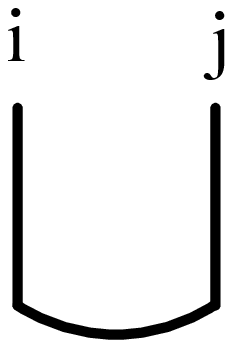}}}

\newcommand{\Nij}{\raisebox{-0.4\height}{\includegraphics[height=0.7cm]{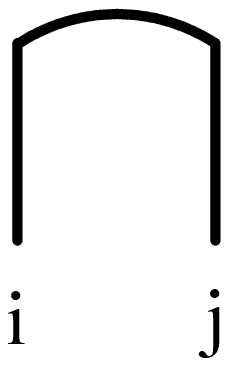}}}
\newcommand{\UN}{\raisebox{-0.4\height}{\includegraphics[height=0.7cm]{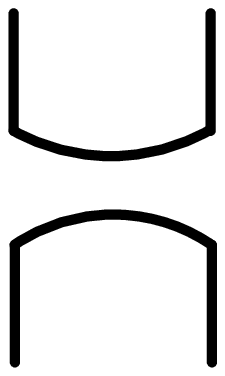}}}
\newcommand{\UNij}{\raisebox{-0.4\height}{\includegraphics[height=0.7cm]{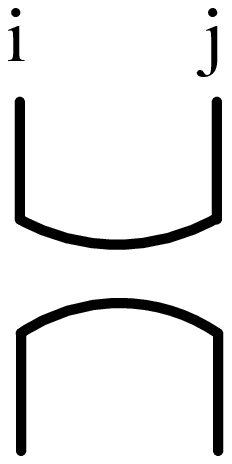}}}
\newcommand{\UU}{\raisebox{-0.4\height}{\includegraphics[height=0.5cm]{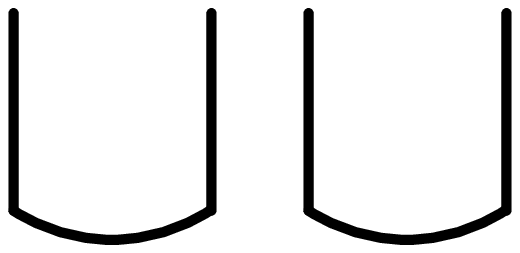}}}
\newcommand{\NN}{\raisebox{-0.4\height}{\includegraphics[height=0.5cm]{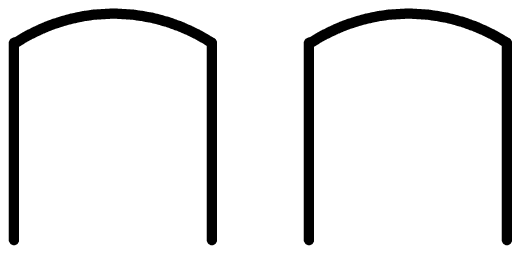}}}
\newcommand{\UinU}{\raisebox{-0.4\height}{\includegraphics[height=0.5cm]{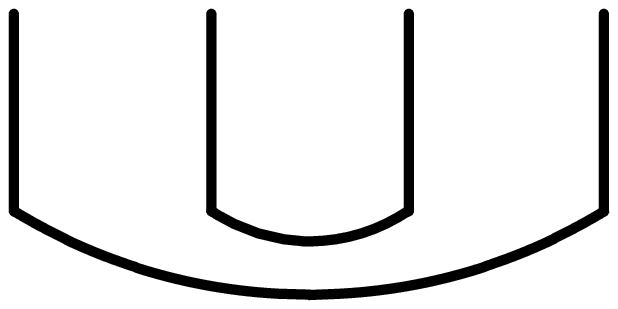}}}
\newcommand{\NinN}{\raisebox{-0.4\height}{\includegraphics[height=0.5cm]{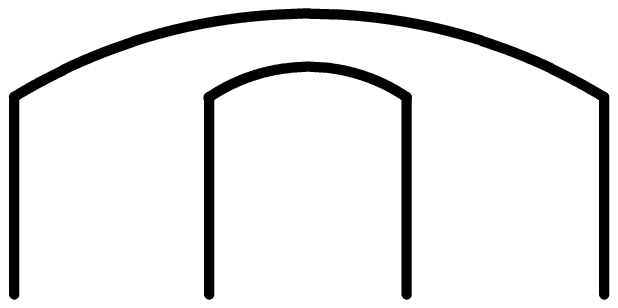}}}
\newcommand{\USU}{\raisebox{-0.4\height}{\includegraphics[height=0.5cm]{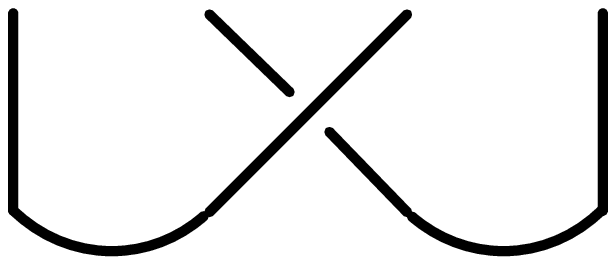}}}
\newcommand{\USSU}{\raisebox{-0.4\height}{\includegraphics[height=0.5cm]{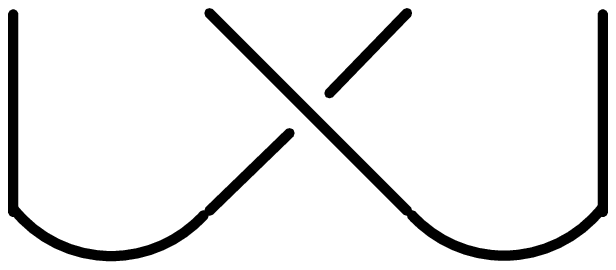}}}
\newcommand{\NSN}{\raisebox{-0.4\height}{\includegraphics[height=0.5cm]{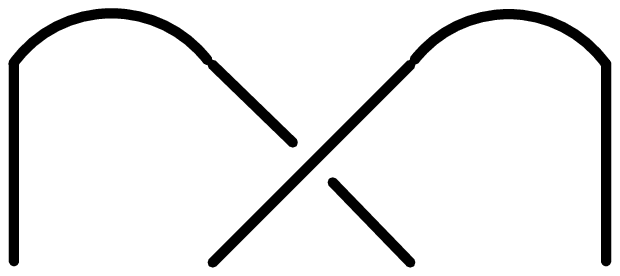}}}
\newcommand{\s}{\raisebox{-0.4\height}{\includegraphics[height=0.5cm]{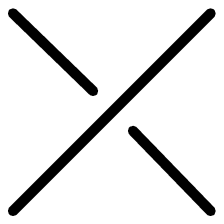}}}
\newcommand{\si}{\raisebox{-0.4\height}{\includegraphics[height=0.5cm]{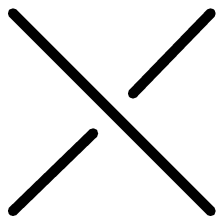}}}
\newcommand{\SU}{\raisebox{-0.4\height}{\includegraphics[height=0.7cm]{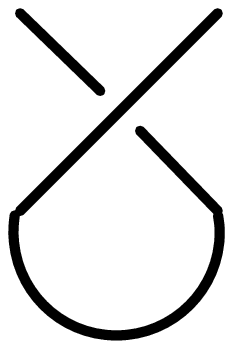}}}
\newcommand{\SiU}{\raisebox{-0.4\height}{\includegraphics[height=0.7cm]{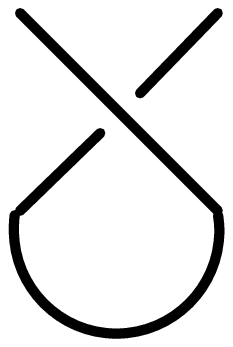}}}
\newcommand{\SUSU}{\raisebox{-0.4\height}{\includegraphics[height=0.7cm]{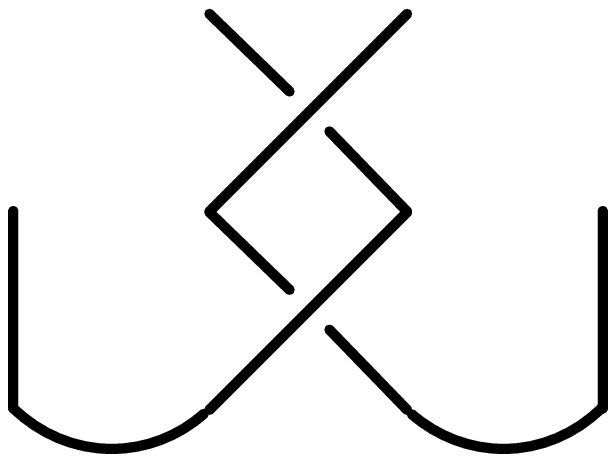}}}
\newcommand{\SUinU}{\raisebox{-0.4\height}{\includegraphics[height=0.7cm]{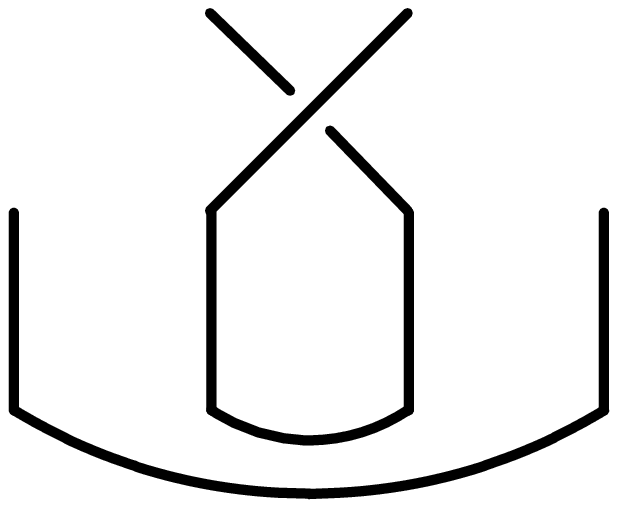}}}
\newcommand{\SUU}{\raisebox{-0.4\height}{\includegraphics[height=0.7cm]{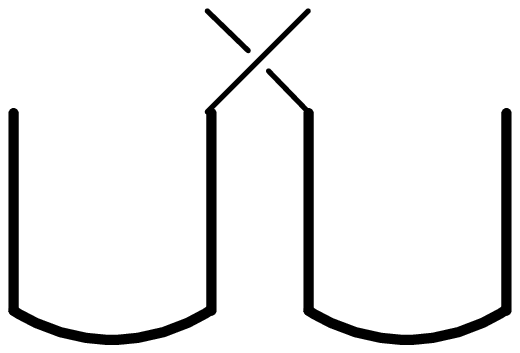}}}
\newcommand{\SiUSU}{\raisebox{-0.4\height}{\includegraphics[height=0.7cm]{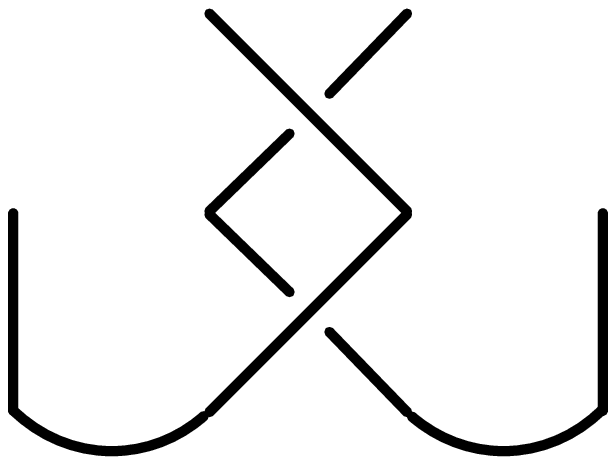}}}
\newcommand{\iiUSU}{\raisebox{-0.4\height}{\includegraphics[height=0.7cm]{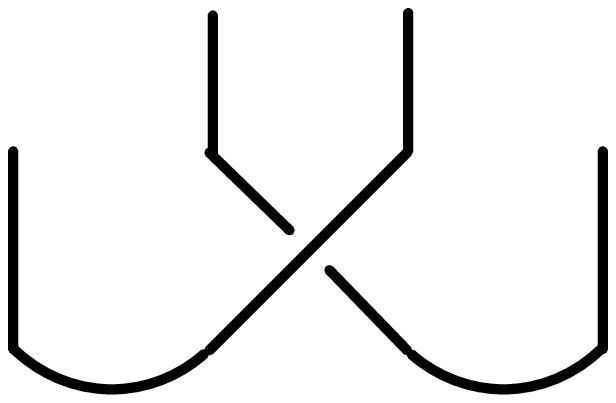}}}
\newcommand{\UNUSU}{\raisebox{-0.4\height}{\includegraphics[height=0.7cm]{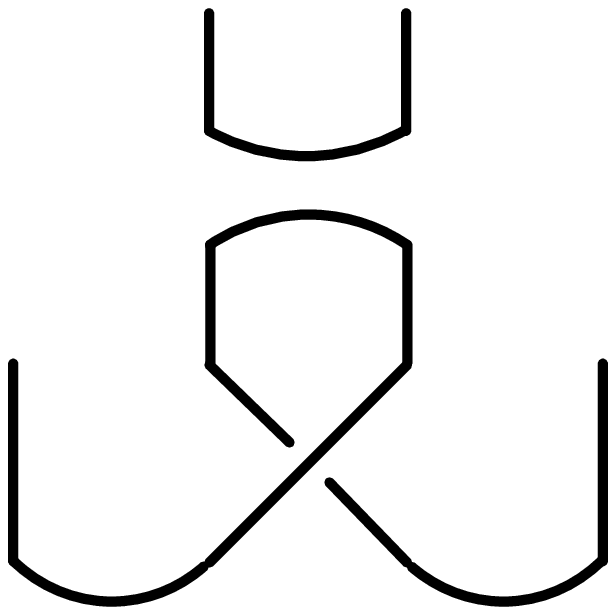}}}

\begin{document}

\title{Birman-Wenzl-Murakami Algebra and the Topological Basis}

\author{Zhou Chengcheng}
\email{Zhoucc237@nenu.edu.cn}
\address{School of Physics, Northeast Normal University,
Changchun 130024, People's Republic of China}
\author{Xue Kang}
\email{Xuekang@nenu.edu.cn}
\address{School of Physics, Northeast Normal University,
Changchun 130024, People's Republic of China}
\author{Wang Gangcheng}
\address{School of Physics, Northeast Normal University,
Changchun 130024, People's Republic of China}
\author{Sun Chunfang}
\address{School of Physics, Northeast Normal University,
Changchun 130024, People's Republic of China}
\author{Du Guijiao}
\address{School of Physics, Northeast Normal University,
Changchun 130024, People's Republic of China}

\begin{abstract}
In this paper, we use entangled states to construct $9\times9$-matrix representations of Temperley-Lieb algebra (TLA), then a family of $9\times9$-matrix representations of Birman-Wenzl-Murakami algebra (BWMA) have been presented. Based on which, three topological basis states have been found. And we apply topological basis states to recast nine-dimensional BWMA into its three-dimensional counterpart. Finally, we find the topological basis states are spin singlet states in special case.
\end{abstract}

 \maketitle

\vspace{0.3cm}

\section{Introduction}
Quantum entanglement (QE) is the most surprising nonclassical property of quantum systems which plays a key role in quantum information and quantum computation processing\cite{ent1,ent2,ent3,ent4}. Because of these applications, QE has become one of the most fascinating topics in quantum information and quantum computation. To the best of our knowledge, the Yang-Baxter equation (YBE) plays an important role in quantum integrable problem, which was originated in solving the one-dimensional \(\delta\)-interacting models\cite{ye} and the statistical models\cite{be}. Braid group representations (BGRs) can be obtained from YBE by giving a particular spectral parameter. BGRs of two and three eigenvalues have direct relationship with Temperley-Lieb algebra (TLA)\cite{tla} and Birman-Wenzl-Murakami algebra (BWMA)\cite{bwma} respectively. TLA and BWMA have been widely used to construct the solutions of YBE\cite{cgx,ybr1,ybr2,ybr3}.

The TLA first appeared in statistical mechanics as a tool to analyze various interrelated lattice models\cite{tla} and was related to link and knot invariants\cite{tla1}. In the subsequent developments TLA is related to knot theory, topological quantum field theory, statistical physics, quantum teleportation, entangle swapping and  universal quantum computation\cite{kk,tla2}. On the other hand, the BWMA\cite{bwma} including braid algebra and TLA was first defined and independently studied by Birman, Wenzl and Murakami. It was designed partially help to understand Kauffman's polynomial in knot theory. Recently, Ref.\cite{hxg} applied topological basis states for spin-1/2 system to recast 4-dimensional YBE into its 2-dimensional counterpart. As we know, few studies have reported topological basis states for spin-1 system. The motivation for our works is to find topological basis states for spin-1 system and study the topological basis states.

The purpose of this paper is twofold: one is that we construct a family of \(9\times 9\)-matrix representations of BWMA; the other concerns topological basis states for spin-1 system. This paper is organized as follows. In Sec. 2, we use entangled states to construct the $9\times9$ matrix representations of TLA, then we present a family of \(9\times 9\)-matrix representations of BWMA, and study the entangled states. In Sec. 3, we obtain three topological basis states of BWMA, and we recast nine-dimensional BWMA into its three-dimensional counterpart. We end with a summary.

\section{\(9\times 9\)-matrix representations of BWMA}

The \(4\times 4\) Hermitian matrix $E$, which satisfies TLA and can construct the well-known six-vertex model\cite{ybei}, takes the representation
\begin{equation}
E=\left(
\begin{array}{cccc}
 0 & 0 & 0 & 0 \\
 0 & q & \eta & 0 \\
 0 & \eta^{-1} & q^{-1} & 0 \\
 0 & 0 & 0 & 0
\end{array}
\right),
\end{equation}

where $\eta=e^{i\varphi}$ with $\varphi$ being any flux. We can rewrite $E$ as that

\begin{equation}
\label{}\left\{ \begin{aligned}
             &E=d|\Psi\rangle\langle\Psi| ,\\
             &|\Psi\rangle=d^{-1/2}(q^{1/2}|\uparrow\downarrow\rangle+q^{-1/2}e^{-i\varphi}|\downarrow\uparrow\rangle),
\end{aligned} \right.
\end{equation}
where $d=q+q^{-1}$.

So like this symmetrical method, we found the \(9\times 9\) Hermitian matrices $E$'s, which satisfies TLA, take the representations as follows
\begin{equation}
\label{}\left\{ \begin{aligned}
             &E=d|\Psi\rangle\langle\Psi| ,\\
             &|\Psi\rangle=d^{-1/2}(q^{1/2}|\lambda\mu\rangle+e^{i\phi_{\nu}}|\nu\nu\rangle+q^{-1/2}e^{i\varphi_{\mu\lambda}}|\mu\lambda\rangle),\\
\end{aligned} \right.
\end{equation}
where $d=q+1+q^{-1}$, $\lambda\neq\mu\neq\nu\in(1,0,-1)$ and $ (d,q,\phi_{\nu},\varphi_{\mu\lambda}) \in real$.
Recently, a \(9\times 9\)--matrix representation of BWMA has been presented\cite{bwma1,bwma2}. We notice that $E$ is the same as Gou \emph{et.al.}\cite{bwma1,bwma2} presented, when $\phi_{\nu}=\varphi_2-\varphi_1+\pi$, $\varphi_{\mu\lambda}=-2\varphi_1$, $\lambda=1$, $\mu=-1$ and $\nu=0$.

As we know the BWMA relations\cite{bwma,cgx,gx2,j} including braid relations and TLA relations satisfy the following relations,
\begin{equation}
\label{bway} \left\{ \begin{aligned}
             &S_i-S_i^{-1}=\omega (I-E_i) ,\\
             &S_iS_{i\pm 1}S_i=S_{i\pm 1}S_iS_{i\pm 1},\ S_iS_j=S_jS_i,|i-j|\geq 2 ,\\
             &E_iE_{i\pm 1}E_i=E_i ,\ E_iE_j=E_jE_i ,\ |i-j|\geq 2 ,\\
             &E_iS_i=S_iE_i=\sigma E_i ,\\
             &S_{i\pm 1}S_iE_{i\pm 1}=E_iS_{i\pm 1}S_i=E_iE_{i\pm 1},\\
             &S_{i\pm 1}E_iS_{i\pm 1}=S_i^{-1}E_{i\pm 1}S_i^{-1} ,\\
             &E_{i\pm 1}E_iS_{i\pm 1}=E_{i\pm 1}S_i^{-1} ,\ S_{i\pm 1}E_iE_{i\pm 1}=S_i^{-1}E_{i\pm 1} ,\\
             &E_iS_{i\pm 1}E_i=\sigma ^{-1}E_i ,\\
             &E_i^2=\left(1-\frac{\sigma-\sigma^{-1}}{\omega}\right)E_i,
             \end{aligned} \right.
\end{equation}
where \(S_i,S_{i\pm 1}\) satisfy the braid relations, \(E_i,E_{i\pm 1}\) satisfy the TLA relations \cite{tla}
\begin{equation}
\label{tla1}\left\{ \begin{aligned}
             &E_iE_{i\pm 1}E_i=E_i ,\ E_iE_j=E_jE_i ,\ |i-j|\geq 2 ,\\
             &E_i^2=d E_i,
\end{aligned} \right.
\end{equation}
where \(0\ne d\in \mathbb{C}\) is topological parameter in the knot theory which does not depend on the sites of lattices. We denote $\sigma=q^{-2}$ and $\omega=q-q^{-1}$ throughout the text. The notations $E_i\equiv E_{i,i+1}$ and $S_i\equiv S_{i,i+1}$ are used, $E_{i,i+1}$ and $S_{i,i+1}$ are abbreviation of $I_1\otimes...\otimes I_{i-1}\otimes E_{i,i+1}\otimes I_{i+2}\otimes...\otimes I_{N}$ and $I_1\otimes...\otimes I_{i-1}\otimes S_{i,i+1}\otimes I_{i+2}\otimes...\otimes I_{N}$ respectively, and $I_j$ represents the unit matrix of the $j$-th particle.

Following the matrix representation of TLA we obtain a family of \(9\times 9\)-matrix representations of BWMA as follows
\begin{equation}
\label{}\left\{ \begin{aligned}
             &E=d|\Psi\rangle\langle \Psi| ,\\
             &|\Psi\rangle=d^{-1/2}(q^{1/2}|\lambda\mu\rangle+e^{i\phi_{\nu}}|\nu\nu\rangle+q^{-1/2}e^{i\varphi_{\mu\lambda}}|\mu\lambda\rangle),
\end{aligned} \right.
\end{equation}

\begin{equation}
\label{} \begin{aligned}
             S&=q(|\lambda\lambda\rangle\langle\lambda\lambda|+|\mu\mu\rangle\langle\mu\mu|)+|\nu\nu\rangle\langle\nu\nu|\\
             &+(q-q^{-1})(|\nu\lambda\rangle\langle\nu\lambda|+|\mu\nu\rangle\langle\mu\nu|)+(q-1)^2(q+1)q^{-2}|\mu\lambda\rangle\langle\mu\lambda|\\
             &+e^{-i\varphi_{\mu\lambda}/2}(|\lambda\nu\rangle\langle\nu\lambda|+|\nu\mu\rangle\langle\mu\nu|)
             +e^{i\varphi_{\mu\lambda}/2}(|\nu\lambda\rangle\langle\lambda\nu|+|\mu\nu\rangle\langle\nu\mu|)\\
             &+q^{-1}e^{-i\varphi_{\mu\lambda}}|\lambda\mu\rangle\langle\mu\lambda|+q^{-1}e^{i\varphi_{\mu\lambda}}|\mu\lambda\rangle\langle\lambda\mu|\\
             &-q^{-3/2}(q^2-1)(e^{i(\phi_{\nu}-\varphi_{\mu\lambda})}|\nu\nu\rangle\langle\mu\lambda|+
             e^{-i(\phi_{\nu}-\varphi_{\mu\lambda})}|\mu\lambda\rangle\langle\nu\nu|),
\end{aligned}
\end{equation}

\begin{equation}
\label{} \begin{aligned}
             S^{-1}&=q^{-1}(|\lambda\lambda\rangle\langle\lambda\lambda|+|\mu\mu\rangle\langle\mu\mu|)+|\nu\nu\rangle\langle\nu\nu|\\
             &+(q^{-1}-q)(|\lambda\nu\rangle\langle\lambda\nu|+|\nu\mu\rangle\langle\nu\mu|)+(q-1)^2(q+1)q^{-1}|\lambda\mu\rangle\langle\lambda\mu|\\
             &+e^{-i\varphi_{\mu\lambda}/2}(|\lambda\nu\rangle\langle\nu\lambda|+|\nu\mu\rangle\langle\mu\nu|)
             +e^{i\varphi_{\mu\lambda}/2}(|\nu\lambda\rangle\langle\lambda\nu|+|\mu\nu\rangle\langle\nu\mu|)\\
             &+qe^{-i\varphi_{\mu\lambda}}|\lambda\mu\rangle\langle\mu\lambda|+qe^{i\varphi_{\mu\lambda}}|\mu\lambda\rangle\langle\lambda\mu|\\
             &+q^{-1/2}(q^2-1)(e^{-i\phi_{\nu}}|\lambda\mu\rangle\langle\nu\nu|+
             e^{i\phi_{\nu}}|\nu\nu\rangle\langle\lambda\mu|),
\end{aligned}
\end{equation}
where $d=q+1+q^{-1}$, $\lambda\neq\mu\neq\nu\in(1,0,-1)$ and $ (d,q,\phi_{\nu},\varphi_{\mu\lambda}) \in real$.

\begin{figure}
\includegraphics[width=10cm]{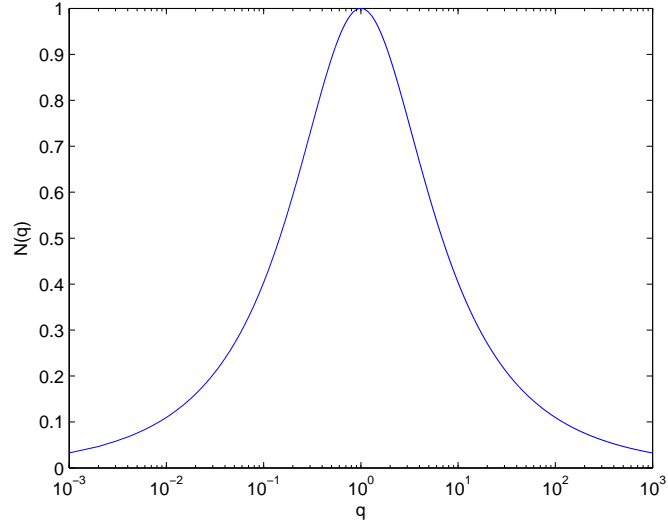}
\caption{(The negativity $\mathcal{N}(q)$ versus the parameter $q$.)}
\label{fig:neg}
\end{figure}

It is worth noticing that the states $|\Psi\rangle$'s are entangled states. By means of negativity, we study these entangled states. The negativity for two
qutrits is given by,
\begin{equation}
\label{}
\mathcal{N}(\rho)\equiv \frac{\parallel\rho^{T_A}\parallel-1}{2},
\end{equation}
where $\parallel\rho^{T_A}\parallel$ denotes the trace norm of $\rho^{T_A}$, which denotes the partial transpose of
the bipartite state $\rho$ \cite{neg}. In fact, $\mathcal{N}(\rho)$ corresponds to the absolute value of the sum of negative eigenvalues of $\rho^{T_A}$ , and negativity vanishes for unentangled states\cite{ent}. By calculation, we can obtain the negativity of states $|\Psi\rangle$'s as
\begin{equation}
\label{}
\mathcal{N}(q)= \frac{q^{1/2}+1+q^{-1/2}}{d},
\end{equation}
where $d=q+1+q^{-1}$. The Fig.\ref{fig:neg} corresponds to the negativity $\mathcal{N}(q)$. One demonstrates that the states $|\Psi\rangle$'s become maximally entangled states of two qutrits as $|\Psi\rangle=(|\lambda\mu\rangle+e^{i\phi_{\nu}}|\nu\nu\rangle+e^{i\varphi_{\mu\lambda}}|\mu\lambda\rangle)/\sqrt{3}$ when $q=1$.

\section{Topological Basis states}
In the topological quantum computation theory, the two-dimensional (2D) braid behavior under
the exchange of anyons\cite{fw} has been investigated based on the $\nu=5/2$ fractional quantum Hall
effect (FQHE)\cite{as}. The orthogonal topological basis states read\cite{as}
\begin{equation}
\label{TB1} \begin{aligned}
|e_1>&=\frac{1}{d}\UU,\\
|e_2>&=\frac{1}{\sqrt{d^2-1}}(\UinU-\frac{1}{d}\UU),
\end{aligned}
\end{equation}
where the parameter $d$ represents the values of a unknotted loop. In Eq.\eqref{TB1} there are two topological graphics $\UU$ and $\UinU$.
For four lattices, we can easy find four graphics $\UU$, $\UinU$, $\USU$, $\USSU$. If we use Skein relations $\s=q^{1/2}\ii+q^{-1/2}\UN$ ($S=q^{1/2}I+q^{-1/2}E$) and $\si=q^{-1/2}\ii+q^{1/2}\UN$ ($S^{-1}=q^{-1/2}I+q^{1/2}E$), where the unknotted loop $d=\D=-q-q^{-1}$, the third and the fourth graphics recast to $\UU$ and $\UinU$ ($\USU=q^{1/2}\UU+q^{-1/2}\UinU$, $\USSU=q^{-1/2}\UU+q^{1/2}\UinU$). So the topological basis states\eqref{TB1} are self-consistent. But in this paper, we focus on BWMA, the braid group representations ($S$) is independent of TLA representations ($E$), and in BWMA $S-S^{-1}=\omega(I-E)$. So we know the graphics $\USU$ and $\USSU$ have one independent graphic. We choose three independent graphics as $\UU$, $\UinU$ and $\USU$.

We define
\begin{equation}
\label{} \begin{aligned}
&\Uij=d^{1/2}|\Phi_{ij}\rangle=q^{1/2}|\lambda\mu\rangle+e^{i\phi_{\nu}}|\nu\nu\rangle+q^{-1/2}|\mu\lambda\rangle,\\
&\Nij=d^{1/2}\langle\Phi_{ij}|=q^{1/2}\langle\lambda\mu|+e^{-i\phi_{\nu}}\langle\nu\nu|+q^{-1/2}\langle\mu\lambda|,\\
&[~\UU~]^{\dag}=\NN,\\
&[~\UinU~]^{\dag}=\NinN,\\
&[~\USU~]^{\dag}=\NSN.
\end{aligned}
\end{equation}
So $E$ recasts to $E_{ij}=\UNij$.
Following the BWMA, we define the graphic rules
\begin{equation}
\label{} \begin{aligned}
&\SU=\sigma \U, \SiU=\sigma^{-1}\U\\
&\s-\si=\omega(\ii-\UN),\\
&\D=d ~(the~ unknotted ~loop).
\end{aligned}
\end{equation}

The orthogonal basis states read
\begin{equation}
\label{}\left\{ \begin{aligned}
             &|e_1\rangle=\frac{q}{(1+q^2)\sqrt{d^2-d-1}}(\USU+q\UinU-\frac{q(q+1)}{d}\UU) ,\\
             &|e_2\rangle=\frac{1}{d}\UU ,\\
             &|e_3\rangle=\frac{q}{(1+q^2)\sqrt{d}}(\USU-q^{-1}\UinU-\frac{q^2-q^{-1})}{d}\UU) .
\end{aligned} \right.
\end{equation}
Let's introduce the reduced operators $E_{A},E_{B},A$ and $B$
\begin{equation}
\label{}\left\{ \begin{aligned}
             &(E_{A})_{ij}=\langle e_i|E_{12}|e_j\rangle ,\\
             &(E_{B})_{ij}=\langle e_i|E_{23}|e_j\rangle ,\\
             &A_{ij}=\langle e_i|S_{12}|e_j\rangle ,\\
             &B_{ij}=\langle e_i|S_{23}|e_j\rangle .
\end{aligned} \right.
\end{equation}
Due to the limited length, we only show how $S_{23}$ acts on $|e_3\rangle$ in detail as follows
\begin{equation}
\label{} \begin{aligned}
             S_{23}|e_3\rangle&=\frac{q}{(1+q^2)\sqrt{d}}(\SUSU-q^{-1}\SUinU-\frac{q^2-q^{-1}}{d}\SUU)\\
             &=\frac{q}{(1+q^2)\sqrt{d}}(\SiUSU+\omega(\iiUSU-\UNUSU)-q^{-1}\sigma \UinU-\frac{q^2-q^{-1}}{d}\USU)\\
             &=\frac{q}{(1+q^2)\sqrt{d}}(\UU+\omega(\USU-\sigma\UinU)-q^{-1}\sigma\UinU-\frac{q^2-q^{-1}}{d}\USU)\\
             &=\frac{q}{(1+q^2)\sqrt{d}}((\omega-\frac{q^2-q^{-1}}{d})\USU-(\omega+q^{-1})\sigma\UinU+\UU)\\
             &=-\frac{\sqrt{d^2-d-1}}{q^2\sqrt{d}(d-1)}|e_1\rangle+\frac{q}{\sqrt{d}}|e_2\rangle+\frac{d-2}{d-1}|e_3\rangle.
\end{aligned}
\end{equation}

Thus their matrix representations in the basis states (\(|e_1\rangle,|e_2\rangle,|e_3\rangle\)) are
given by
\begin{equation}
E_A=diag\{0,d,0\},
\end{equation}

\begin{equation}
E_B=\left(
\begin{array}{>{\displaystyle}l>{\displaystyle}c>{\displaystyle}r}
 \frac{d^2-d-1}{d } & \frac{\sqrt{d^2-d-1} }{d } & -\frac{\sqrt{d^2-d-1} }{\sqrt{d} } \\
 \frac{\sqrt{d^2-d-1} }{d} & \frac{1}{d} & -\frac{1}{\sqrt{d}} \\
 -\frac{\sqrt{d^2-d-1} }{\sqrt{d}} & -\frac{1}{\sqrt{d}} & 1
\end{array}
\right),
\end{equation}

\begin{equation}
A=diag\{q,q^{-2},-q^{-1}\},
\end{equation}

\begin{equation}
B=\left(
\begin{array}{>{\displaystyle}l>{\displaystyle}c>{\displaystyle}r}
 \frac{1}{q^4(d-1)d} & \frac{\sqrt{d^2-d-1} }{d q} & -\frac{\sqrt{d^2-d-1} }{q^2(d-1) \sqrt{d}} \\
 \frac{\sqrt{d^2-d-1} }{d q} & \frac{q^2}{d} & \frac{q}{\sqrt{d}} \\
 -\frac{\sqrt{d^2-d-1} }{q^2(d-1) \sqrt{d}} & \frac{q}{\sqrt{d}} & \frac{d-2}{d-1}
\end{array}
\right),
\end{equation}
where $E_A$, $E_B$, $A$ and $B$ are Hermitian matrices.
It is worth noting that \(E_B=UE_AU^{-1},B=UAU^{-1}\),
\begin{equation}
U=\left(
\begin{array}{>{\displaystyle}l>{\displaystyle}c>{\displaystyle}r}
 \frac{1}{(d-1)d } & -\frac{\sqrt{d^2-d-1}}{d}  & -\frac{ \sqrt{d^2-d-1} }{\sqrt{d} (d-1)} \\
 \frac{\sqrt{d^2-d-1} }{d } & -\frac{1}{d} & \frac{1}{\sqrt{d}} \\
 \frac{ \sqrt{d^2-d-1} }{\sqrt{d} (d-1)} & \frac{1}{\sqrt{d}}& -\frac{d-2}{d-1}
\end{array}
\right),
\end{equation}
and they satisfy the reduced BWMA relations
\begin{equation}
\label{bwa} \left\{ \begin{aligned}
             &A-A^{-1}=\omega (I-E_A) ,\ B-B^{-1}=\omega (I-E_B),\\
             &ABA=BAB ,\\
             &E_AE_BE_A=E_A ,\ E_BE_AE_B=E_B ,\\
             &E_AA=AE_A=\sigma E_A ,\ E_BB=BE_B=\sigma E_B ,\\
             &ABE_A=E_BAB=E_BE_A ,\ BAE_B=E_ABA=E_AE_B ,\\
             &AE_BA=B^{-1}E_AB^{-1} ,\ BE_AB=A^{-1}E_BA^{-1} ,\\
             &E_AE_BA=E_AB^{-1} ,\ E_BE_AB=E_BA^{-1} ,\\
             &AE_BE_A=B^{-1}E_A ,\ BE_AE_B=A^{-1}E_B ,\\
             &E_ABE_A=\sigma^{-1}E_A ,\ E_BAE_B=\sigma^{-1}E_B, \\
             &E_A^2=(1-\frac{\sigma-\sigma^{-1}}{\omega})E_A ,\
             E_B^2=(1-\frac{\sigma-\sigma^{-1}}{\omega})E_B.
      \end{aligned} \right.
\end{equation}
We emphasize that \eqref{bwa} acts on the basis ($|e_1\rangle,|e_2\rangle,|e_3\rangle$).

It is worth noting that the topological basis states are singlet states, when $\phi_{\nu}=\pi$, $\lambda=1$, $\mu=-1$, $\nu=0$ and $q=1$. In other words, $S^2|e_i\rangle=0$ and $S_z|e_i\rangle=0$, where $S=\sum_1^4S_j$, $S_j$ are the operators of spin-1 angular momentum for the $j$-th particle, $i=1,2,3$.

\section{Summary}
In this paper we construct $9\times9$-matrix representations of TLA, where we used the entangled states ($|\Psi\rangle=d^{-1/2}(q^{1/2}|\lambda\mu\rangle+e^{i\phi_{\nu}}|\nu\nu\rangle+q^{-1/2}e^{i\varphi_{\mu\lambda}}|\mu\lambda\rangle)$). Then we get a family of $9\times9$ representations of BWMA. We study the entangled states $|\Psi\rangle$'s, and find the negativity related parameter $q$. The negativity became the maximum value if $q=1$. In Sce. 3, we defined the third topological graphic $\USU$ and find three orthogonal topological basis states of BWMA, based on the former researchers. It was mentioned that the Hermitian matrices $E_A$, $E_B$, $A$ and $B$ have an interesting similar transformation matrix $U$ which satisfies $B=UAU^{-1}$ and $E_B=UE_AU^{-1}$. Based on them, we obtain a three-dimensional representation of BWMA. Finally we find the topological basis states are the spin singlet states, if $\phi_{\nu}=\pi$, $\lambda=1$, $\mu=-1$, $\nu=0$ and $q=1$. Our next work will study how the topological basis states play a role in quantum theory.

\section{Acknowledgments}
This work was supported by NSF of China (Grant No.10875026)

\end{document}